\documentclass[12pt,journal,draftclsnofoot,onecolumn]{IEEEtran}
\usepackage{amsfonts}
\usepackage{amssymb}
\usepackage{cite}
\usepackage[cmex10]{amsmath}
\usepackage{float}
\usepackage{color}
\usepackage{stfloats,fancyhdr}
\usepackage{amsmath}
\usepackage{amsthm}
\usepackage{algorithm}
\usepackage{algorithmic}
\usepackage{multirow}
\usepackage{changepage}
\usepackage[normalem]{ulem}

\usepackage{url}

\newtheorem{definition}{Definition}

\newtheorem{remark}{Remark}

\IEEEoverridecommandlockouts

\ifCLASSINFOpdf
  \usepackage[pdftex]{graphicx}
  \DeclareGraphicsExtensions{.pdf,.jpeg,.png}
\else
  \usepackage[dvips]{graphicx}
  \DeclareGraphicsExtensions{.pdf}
\fi

\usepackage{subfigure}

\usepackage{fancybox,dashbox}

\begin{document}

\title{Secure Status Updates under Eavesdropping: Age of Information-based Physical Layer Security Metrics}

\author{He Chen$^*$, Qian Wang$^*$, Parthajit Mohapatra, and Nikolaos Pappas
\thanks{$^*$The first two authors contributed equally to this work. The work of H. Chen is supported by the CUHK direct grant under the project code 4055126.

H. Chen is with the Department of Information Engineering, The Chinese University of Hong Kong, Hong Kong SAR, China (Email: he.chen@ie.cuhk.edu.hk).

Q. Wang is with the School of Electrical and Information Engineering, The University of Sydney, NSW 2006, Australia (Email: qian.wang2@sydney.edu.au). The work of Q. Wang is done when she is a visiting student at The Chinese University of Hong Kong.

P. Mohapatra is with Department of Electrical Engineering, Indian Institute of Technology (IIT) Tirupati, Indian (Email: parthajit@iittp.ac.in).

N. Pappas is with the Department of Science and Technology, Link\"{o}ping University, SE-60174 Norrk\"{o}ping, Sweden (Email:
nikolaos.pappas@liu.se).

}}

\maketitle

\begin{abstract}
This letter studies the problem of maintaining information freshness under passive eavesdropping attacks. The classical three-node wiretap channel model is considered, in which a source aims to send its latest status wirelessly to its intended destination, while protecting the message from being overheard by an eavesdropper. Considering that conventional channel capacity-based secrecy metrics are no longer adequate to measure the information timeliness in status update systems, we define two new age of information-based metrics to characterize the secrecy performance of the considered system. We further propose, analyze, and optimize a randomized stationary transmission policy implemented at the source for further enhancing the secrecy performance. Simulation results are provided to validate our analysis and optimization.
\end{abstract}


%
\IEEEpeerreviewmaketitle

\section{Introduction}
With the quick proliferation of the Internet of Things (IoT) technologies, more and more devices and equipment are connected to the Internet. This brings severe security and privacy concerns considering that a large share of devices will be served by wireless communication technologies \cite{sicari2015security,Ericsson2016}. Conventionally, various cryptography techniques have been invented and applied at higher layers to protect the security of computer networks. However, these approaches normally involve intensive computation, which could be time- and energy-consuming for low-cost and low-complexity IoT devices. In this context, physical layer security techniques that leverage the properties of wireless physical channels, such as interference and fading, to further strengthen the security of wireless communication systems, have been regarded as appealing complements or even alternative solutions \cite{Mukherjee2015pieee}.

The research on physical layer security was pioneered by Wyner in his seminal work \cite{wyner1975wire}, where he proved that perfect secrecy can be realized when the channel from source to eavesdropper is a degraded version of that from source to destination. Since then, enormous efforts have been put to confound eavesdropping from either the perspective of information theory (see \cite{bloch2011physical} and references therein) or signal processing (see \cite{mukherjee2014principles} for a comprehensive literature survey). These studies were mainly built on the secrecy performance metrics like secrecy capacity and secrecy outage probability, that are defined based on the information-theoretic channel capacity concept.

On the other hand, most IoT applications involve remote monitoring, where sensors are deployed to monitor various physical processes, for example, the air pollution level and the health conditions of running machines in a factory. The sensed information is then reported to a remote entity like a cloudlet for further analysis. In these contexts, the information generally has higher value when it is fresher. This represents a strong motivation for a proper characterization of the information freshness. Recent results showed that conventional network metrics like throughput or delay are no longer appropriate for quantifying the information freshness. A new metric, termed age of information (AoI), has been coined in \cite{Kaul2011mini} for evaluating the information freshness, which have attracted increasing attentions very recently \cite{kosta2017age,sun2019age,Gu2019timely,Abd2019on,chen2020age}. The AoI is defined as the time elapsed since the generation time of the last status update received at the destination. In this context, a natural and fundamental question that arises is ``\emph{how to quantify the secrecy performance of status update systems from the perspective of physical layer}?" To the best of our knowledge, this is still an open problem in the literature and this letter serves as the first attempt to answer it. It is worth mentioning here a handful of studies that investigated the problem of maintaining information freshness under active jamming attacks, see e.g., \cite{xiao2018dynamic,garnaev2019maintaining}, which is fundamentally different from the passive eavesdropping attacks considered in this letter.

In this letter, we consider a status update system consisting of a source, a destination, and an eavesdropper. The source aims to report its latest status to the destination under the risk of being wiretapped by the eavesdropper. From a system security perspective, the source pursues the information at the destination to be much fresher than that at the eavesdropper. To characterize the secrecy performance of such a status update system, we define two new secrecy metrics drawing on the AoI concept, namely secrecy age and secrecy age outage probability. We further propose, analyze, and optimize a randomized stationary transmission policy adopted at the source for enhancing the system secrecy performance. Simulation results are carried out to affirm our analysis and design.


\section{System Model and New Secrecy Metrics}
Consider a status update system consisting of a source ($S$), a destination ($D$), and an eavesdropper ($E$). Specifically, $S$ aims to report its latest status update to $D$ via an error-prone wireless channel, while the passive adversary $E$ attempts to eavesdrop the transmission of $S$ to grasp its latest status\footnote{Note that successful eavesdropping is the first step towards launching other active attacks like replay attacks.}. We assume slotted time, time is divided into slots of equal duration, and the transmission of a status updates occupies one slot. The timeliness and freshness of status updates at the receivers including $D$ and $E$ is measured by the AoI metric.

The \emph{generate-at-will} status generation model is considered. That is, a new status update is generated at $S$ at the beginning of each time slot. However, $S$ may choose to drop the newly generated status according to different transmission policies to be elaborated later in this section. In this letter, we investigate a challenging scenario, where $S$ has no knowledge of the channel conditions from itself to both $D$ and $E$. Instead, it only knows the probabilities of its status update transmissions to be successfully received by $D$ and $E$, denoted by $p$ and $q$, respectively. Here, to make it general, we choose not to specify a particular channel fading model for characterizing the successful transmission probabilities. This is because once the channel fading model is given, we can link $p$ and $q$ with the average signal-to-noise ratios of the corresponding links.

We use $I_D(t)$ and $I_E(t)$ to denote the transmission success indicators of the $S-D$ and $S-E$, respectively. Specifically, $I_X (t) = 1$, $X\in \{D,E\}$, if the status update over the $S-X$ link is successful during time slot $t$, and $I_X (t) = 0$ otherwise. Denote by $\delta_D(t)$ and $\delta_E(t)$ the instantaneous ages at $D$ and $E$ in the $t$-th time slot, respectively. The instantaneous ages of these two receiving nodes in the time slot $t+1$ can then be expressed as
\begin{equation}
\delta _{X} \left( {t  + 1} \right) =\left\{
\begin{matrix}
\begin{split}
   &{1}, \quad\text{if}~ S~{\rm transmits},  {~\rm and~} I_X \left( t  \right)=1, \\
   &{\delta _X \left( {t } \right)+1}, \quad\text{otherwise},  \\
\end{split}
\end{matrix}
\right.
\end{equation}
where $X\in \{D,E\}$.

\subsection{New Secrecy Metrics}
In this letter, we are interested in protecting the status update system against eavesdropping attacks from the perspective of physical layer, which corresponds to the extreme adversary case with no cryptography techniques implemented at higher layers. In this context, $S$ can only resort to the transmission policy to enhance the system secrecy performance. For conventional channel capacity-based physical layer security metrics, such as secrecy capacity and secrecy outage probability, have been defined and widely investigated for various network setups, see e.g., the comprehensive survey papers \cite{bloch2011physical,mukherjee2014principles} and references therein. More specifically, denote by $C_D$ and $C_E$ as the instantaneous channel capacity of the $S-D$ and $S-E$ links, respectively. The secrecy capacity and secrecy outage probability can then be defined as $C_s = \left[C_D-C_E\right]^+$ and $\Pr\left(C_s < R_s\right)$, respectively, with $R_s$ representing the target secrecy rate. However, these conventional secrecy metrics are no longer applicable to status update systems as they are inadequate to quantify the information freshness.

As the first attempt to measure the secrecy performance of status update systems, we now define two new secrecy metrics. Intuitively, to enhance the secrecy of the system, we should make the instantaneous age at $E$ larger than that at $D$ as much as possible. That is, the information at $E$ is much staler than that at $D$. Motivated by this intuition and inspired by the conventional channel capacity-based secrecy metrics, we devise two new secrecy metrics defined in the following:
\begin{definition}
We define the instantaneous {secrecy age} $\delta_s$~as
\begin{equation}\label{eq:secrecy_age}
\delta_s = [\delta_E-\delta_D]^+,
\end{equation}
where $[.]^+$ indicates $\max\{\delta_E-\delta_D,0\}$. 
\end{definition}
\begin{definition}
We then define the {secrecy age outage probability}~as
\begin{equation}\label{eq:age_outage}
\mathcal{P}_{\rm out} = \Pr\left([\delta_E-\delta_D]^+ \le \eta_{\rm th}\right),
\end{equation}
where $\eta_{\rm th}$ is the target information lag between $E$ and $D$. 
\end{definition}
According to Definition 1-2, larger secrecy age and lower secrecy age outage probability are more desirable.

\subsection{Transmission Policy}
We now describe the transmission policy proposed in this letter. Recall that we consider the challenging case where $S$ has no knowledge of the instantaneous channel state information of both $S-D$ and $S-E$ links. Furthermore, there is no feedback from $D$, and thus $S$ does not grasp the instantaneous age at $D$ either. For this case, we propose a randomized stationary policy to be implemented at $S$. More specifically, $S$ transmits a newly generated status update at the beginning of each time slot with a fixed probability $p_{\rm tx}$. When $p_{\rm tx} =  1$, $S$ always transmits a new status update in each time slot. Note that no retransmission is considered in our policy since we can generate a new status update in each time slot.



\section{Analysis and Optimization of the Randomized Stationary Policy}

\subsection{Secrecy Performance Analysis}
To evaluate the average secrecy performance of the randomized stationary policy, we apply a two-dimension MC to characterize the state transition of the considered system, in which the system state $\left(m,n\right)$ indicates that the current age at $D$ is $m$ and that at $E$ is $n$. We can readily verify that the considered MC is irreducible and thus it admits a unique steady state distribution. Denote by $\pmb{\pi}$ the steady state distribution of the considered system implementing the randomized stationary policy, with the entry $\pi_{i,j}$ representing the steady probability of the state $\left(i,j\right)$.

We first notice that the system can transit from any state to the state $(1,1)$ when $S$ transmits and the transmission is successfully received by both the legitimate receiver $D$ and the eavesdropper $E$. We thus have the following equation
\begin{equation}
\sum_{j=1}^{\infty}\sum_{i=1}^{\infty}\pi_{i,j}p_{\rm tx}pq=\pi_{1,1}.
\end{equation}
By noting the fact $\sum_{j=1}^{\infty}\sum_{i=1}^{\infty}\pi_{i,j}=1$, we then have $\pi_{1,1}=p_{\rm tx}pq$. Furthermore,
\begin{equation}\label{eq:stationary_probability1}
\sum_{j=1}^{\infty}\sum_{i=1}^{\infty}\pi_{i,j}p_{\rm tx}p=\sum_{j=1}^{\infty}\pi_{1,j}{\bf \Rightarrow} \sum_{j=1}^{\infty}\pi_{1,j}=p_{\rm tx}p.
\end{equation}
\begin{equation}\label{eq:stationary_probability2}
\sum_{j=1}^{\infty}\sum_{i=1}^{\infty}\pi_{i,j}p_{\rm tx}q=\sum_{i=1}^{\infty}\pi_{i,1} \Rightarrow  \sum_{i=1}^{\infty}\pi_{i,1}  = p_{\rm tx}q.
\end{equation}
\begin{equation}\label{eq:stationary_probability3}
\pi_{i+1,1}=\sum_{j=1}^{\infty}\pi_{i,j}p_{\rm tx}(1-p)q,
\end{equation}
\begin{equation}\label{eq:stationary_probability4}
\pi_{1,j+1}=\sum_{i=1}^{\infty}\pi_{i,j}p_{\rm tx}p(1-q).
\end{equation}
\begin{equation}\label{eq:stationary_probability5}
\pi_{i+1,j+1}=\pi_{i,j}\left[p_{\rm tx}(1-q)(1-p)+ 1 - p_{\rm tx}\right].
\end{equation}

\begin{equation}\label{eq:stationary_probability6}
\sum_{j=1}^{\infty}\pi_{i+1,j}=\sum_{j=1}^{\infty}\pi_{i,j}\left[p_{\rm tx}(1-p)+ 1 - p_{\rm tx}\right]
\end{equation}

\begin{equation}\label{eq:stationary_probability7}
\sum_{i=1}^{\infty}\pi_{i,j+1}=\sum_{i=1}^{\infty}\pi_{i,j}\left[p_{\rm tx}(1-q)+ 1 - p_{\rm tx}\right]
\end{equation}
{Combining the equalities in \eqref{eq:stationary_probability1} and \eqref{eq:stationary_probability3}, we have $\pi_{2,1}=p_{\rm tx}pp_{\rm tx}(1-p)q$. We then jointly consider the equalities in \eqref{eq:stationary_probability1} and \eqref{eq:stationary_probability6} and attain
\begin{equation}\label{eq:sum_pi2j}
\sum_{j=1}^{\infty}\pi_{2,j}=p_{\rm tx}p\left[p_{\rm tx}(1-q)+ 1 - p_{\rm tx}\right].
\end{equation}
Then, based on \eqref{eq:sum_pi2j}, \eqref{eq:stationary_probability3} and \eqref{eq:stationary_probability6}, we can obtain the expressions of $\pi_{3,1}$ and $\sum_{j=1}^{\infty}\pi_{3,j}$ given by
\[\pi_{3,1}=p_{\rm tx}(1-p)qp_{\rm tx}p\left[p_{\rm tx}(1-q)+ 1 - p_{\rm tx}\right],\]
\[\sum_{j=1}^{\infty}\pi_{3,j}=p_{\rm tx}p\left[p_{\rm tx}(1-q)+ 1 - p_{\rm tx}\right]^2.\]
By forward induction, we further have
\[\pi_{i,1}=p_{\rm tx}pp_{\rm tx}(1-p)q{\left(p_{\rm tx}(1-p)+1-p_{\rm tx}\right)}^{i-2},\; \forall i\geq 2.\]

Similarly, we obtain $\pi_{1,2}=p_{\rm tx}qp_{\rm tx}(1-q)p$, $\pi_{1,j}=p_{\rm tx}qp_{\rm tx}(1-q)p{\left(p_{\rm tx}(1-q)+1-p_{\rm tx}\right)}^{j-2}$, $\forall j\geq 2$, and

\begin{equation}
\begin{aligned}
\pi_{i,j}=&\pi_{i-j+1,1}{\left[p_{\rm tx}(1-q)(1-p)+ 1 - p_{\rm tx}\right]}^{j-1}\\
=&p_{\rm tx}pp_{\rm tx}(1-p)q{\left(p_{\rm tx}(1-p)+1-p_{\rm tx}\right)}^{i-j-1}\times\\
&{\left[p_{\rm tx}(1-q)(1-p)+ 1 - p_{\rm tx}\right]}^{j-1}, \text{if } i> j,
\end{aligned}
\end{equation}
\begin{equation}
\begin{aligned}
\pi_{i,j}=&\pi_{1,j-i+1}{\left[p_{\rm tx}(1-q)(1-p)+ 1 - p_{\rm tx}\right]}^{i-1}\\
=&p_{\rm tx}qp_{\rm tx}(1-q)p{\left(p_{\rm tx}(1-q)+1-p_{\rm tx}\right)}^{j-i-1}\times\\
&{\left[p_{\rm tx}(1-q)(1-p)+ 1 - p_{\rm tx}\right]}^{i-1}, \text{if } j> i,
\end{aligned}
\end{equation}
\begin{equation}
\begin{aligned}
\pi_{i,i}&=\pi_{1,1}{\left[p_{\rm tx}(1-q)(1-p)+ 1 - p_{\rm tx}\right]}^{i-1}\\
&=p_{\rm tx}qp{\left[p_{\rm tx}(1-q)(1-p)+ 1 - p_{\rm tx}\right]}^{i-1},\text{if } i\geq 1.
\end{aligned}
\end{equation}

We now can derive closed-form expressions for the average secrecy age and the secrecy age outage probability as follows
\begin{equation}\label{eq:averageage}
\begin{aligned}
	&\mathbb{E}\left[ \delta_s \right]= \mathbb{E}\left[[\delta_E-\delta_D]^+\right]=\sum_{i=1}^{\infty}\sum_{j=i+1}^{\infty}\pi_{i,j}(j-i)\\
	=&\sum_{i=1}^{\infty}{\left[p_{\rm tx}(1-q)(1-p)+ 1 - p_{\rm tx}\right]}^{i-1} \times\\
&\sum_{j=i+1}^{\infty}\pi_{1,2}{\left(p_{\rm tx}(1-q)+1-p_{\rm tx}\right)}^{j-i-2}(j-i)\\
	\overset{(a)}{=}&\frac{1}{p_{\rm tx}(p+q-pq)}\frac{p_{\rm tx}qp_{\rm tx}(1-q)p}{{(p_{\rm tx}q)}^2}=\frac{p(1-q)}{p_{\rm tx}q(p+q-pq)}.
\end{aligned}
\end{equation}
\begin{equation}\label{eq:outageexpress}
\begin{aligned}
 &\mathcal{P}_{\rm out} = \Pr\left([\delta_E-\delta_D]^+\leq \eta_{\rm{th}}\right)\\
= &1- P\left([\delta_E-\delta_D]^+ > \eta_{\rm{th}}\right)=1-\sum_{i=1}^{\infty}\sum_{j=i+1+\eta_{\rm{th}}}^{\infty}\pi_{i,j}\\
= &1-\sum_{i=1}^{\infty}{\left[p_{\rm tx}(1-q)(1-p)+ 1 - p_{\rm tx}\right]}^{i-1}\times\\
 &\sum_{j=i+1+\eta_{\rm{th}}}^{\infty}\pi_{1,2}{\left(p_{\rm tx}(1-q)+1-p_{\rm tx}\right)}^{j-i-2}\\
 \overset{(b)}{=}&1-\frac{(1-q)p{\left(1-p_{\rm tx}q\right)}^{\eta_{\rm{th}}-1}}{p+q-pq},
\end{aligned}
\end{equation}
where the equalities (a) and (b) follow by adopting the infinite sum equation given by \cite[Eq. 0.231]{gradshteyn2014table}.
\begin{remark}
Based on the attained closed-from expressions given in \eqref{eq:averageage} and \eqref{eq:outageexpress}, we can readily verify that the average secrecy age $\mathbb{E}\left[ \delta_s \right]$ and the secrecy age outage probability $\mathcal{P}_{\rm out}$ are decreasing and increasing functions of the transmission probability ${p_{{\rm{tx}}}}$, respectively. We can also see from \eqref{eq:outageexpress} that when $q\rightarrow 0$, $\mathcal{P}_{\rm out}\rightarrow 0$.
\end{remark}

\subsection{Optimization of the Transmission Probability}
We now formulate a problem to optimize the transmission probability ${p_{{\rm{tx}}}}$ at $S$. Intuitively, the age at $D$ is statistically smaller if $S$ transmits more frequently with a larger ${p_{{\rm{tx}}}}$, which, on the other hand, will cause a higher secrecy age outage probability (See Remark 1). Motivated by this intuition, we formulate the following optimization problem
\begin{equation}\label{eq:opt_random}
{\rm \bf{Problem~1}}:\;  \mathop {\max }\limits_{{p_{{\rm{tx}}}}\in(0,1]} {p_{{\rm{tx}}}}\left[ {1 - {{\cal P}_{{\rm{out}}}\left(p_{{\rm{tx}}}\right)}} \right].
\end{equation}

To resolve the above optimization problem, we calculate the first-order derivative of the objective function with respect to ${p_{{\rm{tx}}}}$ and set it equal to zero. After some mathematical manipulations, we can obtain that the optimal ${p_{{\rm{tx}}}}$ is given~by
\begin{equation}\label{eq:optimal_ptx}
{p^*_{{\rm{tx}}}} = \min\left\{{1}/({q{\eta_{\rm{th}}}}),1\right\}.
\end{equation}
We note that the optimal transmission probability does not depend on $p$.

\section{Simulations and Discussions}
We now present simulation results to validate the theoretical analysis conducted above and illustrate the impacts of various parameters on the system secrecy performance.

We plot the curves of the average secrecy age versus the ratio $p/q$ for different combinations of $q$ and $p_{\rm{tx}}$ in Fig. \ref{fig1}. We can see from the figure that the simulation results match well with their analytical counterparts, which verifies the correctness of our theoretical analysis in Section III. Moreover, for fixed $q$ and $p_{\rm{tx}}$, the average secrecy age increases as $p$ increases. This is expected since a larger $p$ means a higher successful transmission probability over the $S-D$ link. On the other hand, when $p$ and $p_{\rm{tx}}$ are given, the larger the value of $q$, the smaller the average secrecy age. This is because $E$ can eavesdrop the status updates from $S$ with a higher success probability. We can also observe from Fig. \ref{fig1} that the average secrecy age decreases as the transmission probability $p_{\rm{tx}}$ increases, which coincides with our analysis in Remark 1.

\begin{figure}
	\centering \scalebox{0.8}{\includegraphics{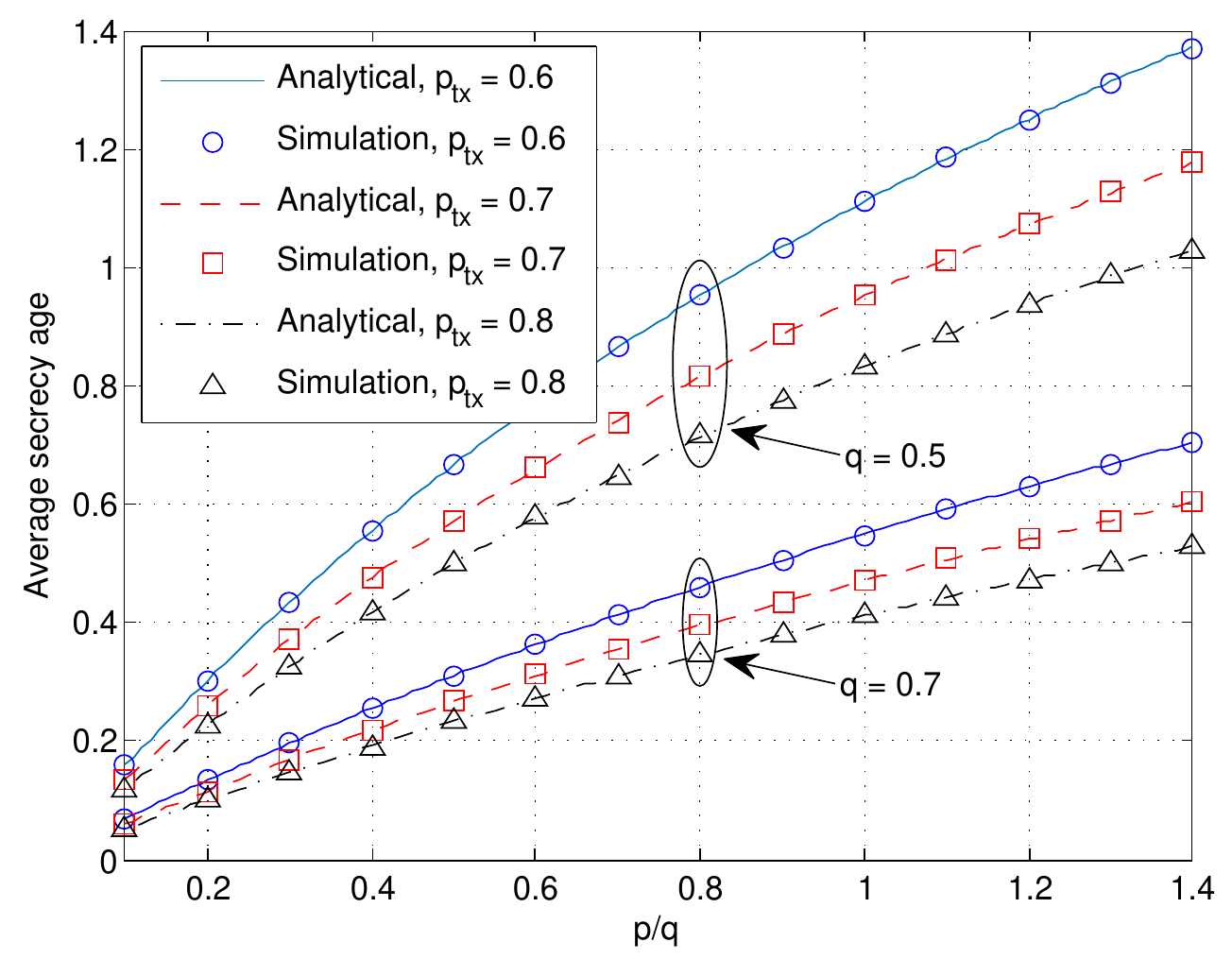}}
	\caption{The average secrecy age versus the the ratio $p/q$ for different values of $q$ and $p_{\rm{tx}}$.}
	\label{fig1}
		\vspace{-1em}
\end{figure}

\begin{figure}
	\centering \scalebox{0.8}{\includegraphics{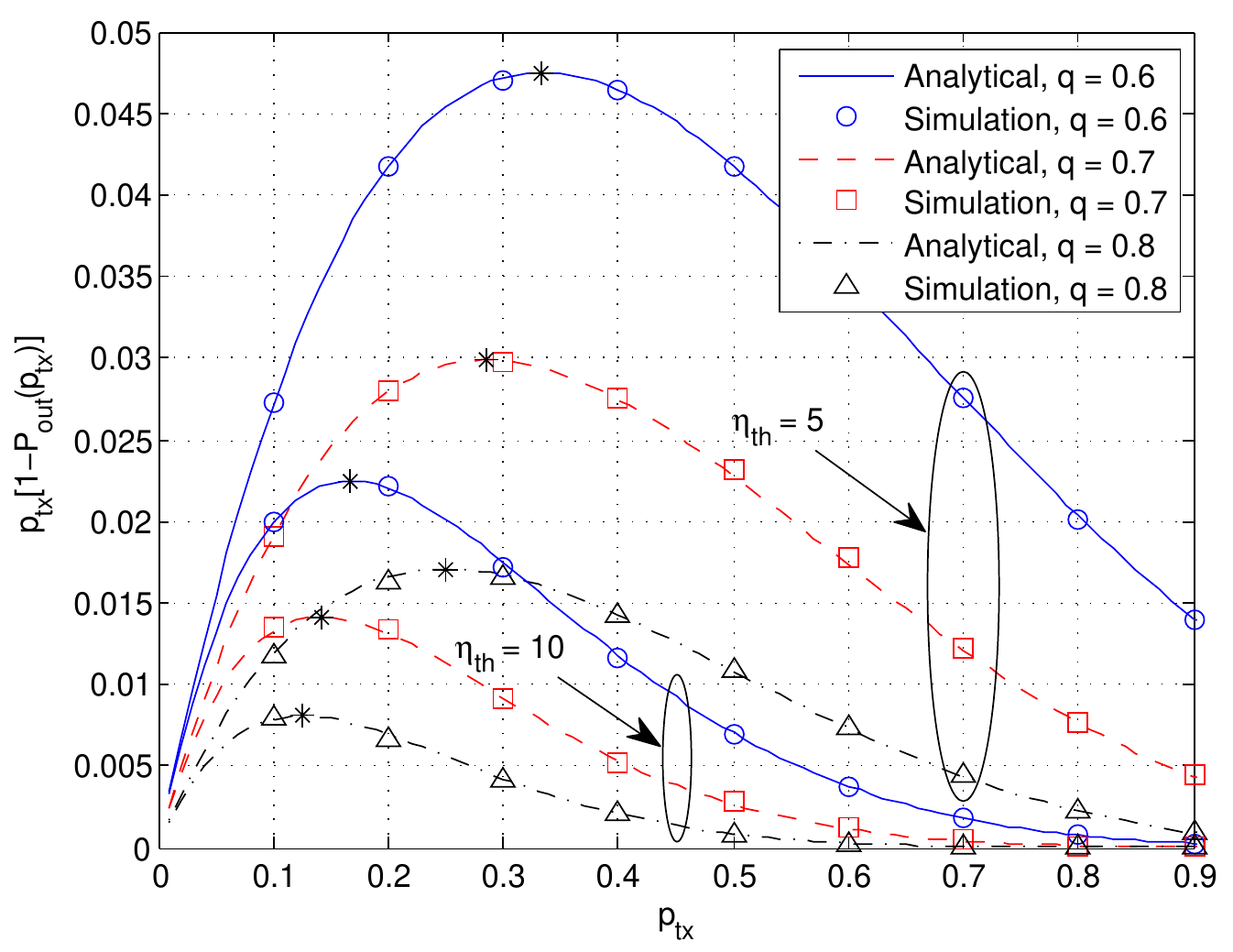}}
	\caption{The curves of the term ${p_{{\rm{tx}}}}\left[ {1 - {{\cal P}_{{\rm{out}}}\left(p_{{\rm{tx}}}\right)}} \right]$, versus the transmission probability ${p_{{\rm{tx}}}}$ with varying secrecy age outage threshold $\eta_{\rm th}$ and $q$, where $p = 0.8$ and the values of ${p^*_{{\rm{tx}}}}\left[ {1 - {{\cal P}_{{\rm{out}}}\left(p^*_{{\rm{tx}}}\right)}} \right]$ corresponding to the optimal solution given in \eqref{eq:optimal_ptx} are labelled by $*$.}
	\label{fig2}
		\vspace{-1em}
\end{figure}

We depict in Fig. \ref{fig2} the curves of the objective function in Problem 1, ${p_{{\rm{tx}}}}\left[ {1 - {{\cal P}_{{\rm{out}}}\left(p_{{\rm{tx}}}\right)}} \right]$, versus the transmission probability ${p_{{\rm{tx}}}}$ with varying secrecy age outage threshold $\eta_{\rm th}$ and $q$. As we can see from the figure, the simulation and analytical results once again coincide. In addition, there is an optimal ${p_{{\rm{tx}}}}$ in all simulated cases. Furthermore, the optimal values corresponding to the closed-form solution to Problem 1 appear at the peaks of the curves, which validates the attained solution. We also can observe from Fig. \ref{fig2} that the optimal value of ${p_{{\rm{tx}}}}$  shifts to the left as either $\eta_{\rm th}$ or $q$ increases, which again affirms the expression of the optimal ${p_{{\rm{tx}}}}$ given by ${p^*_{{\rm{tx}}}} = \min\left\{{1}/({q{\eta_{\rm{th}}}}),1\right\}$.
%
\section{Conclusions}
This letter studied the problem of maintaining information freshness of status update systems under eavesdropping attacks. To that end, we studied the classical three-node wiretap channel. Considering that the conventional channel capacity-based secrecy metrics are no longer adequate to characterize the secrecy performance of status update systems, we proposed two new age of information-based secrecy metrics, termed secrecy age and secrecy age outage probability. We also derived the closed-form expressions of the average secrecy age and the secrecy age outage probability for the considered three-node system by applying a two-dimensional Markov chain to model the dynamic age evolutions at both the destination and the eavesdropper. We further designed and optimized a randomized stationary policy implemented at the source for enhancing the system secrecy performance. Simulation results were finally provided to validate and substantiate our design and analysis.


%
%



%
%

\bibliography{ref}
\bibliographystyle{IEEEtran}
\end{document}